\documentclass[12pt]{article}
\usepackage{bbm}
\usepackage{graphicx}
\usepackage{amsmath}
\usepackage{amssymb}
\usepackage{caption2}
\begin{document}
\bibliographystyle{prsty}
\begin{center}
{\large {\bf \sc{ Analysis of the coupling constants
$g_{a_0\eta\pi^0}$ and  $g_{a_0\eta'\pi^0}$  with light-cone QCD sum rules }}} \\[2mm]
Zhi-Gang Wang \footnote{E-mail,wangzgyiti@yahoo.com.cn. }    \\
 Department of Physics, North China Electric Power University, Baoding 071003, P. R. China \\
\end{center}

\begin{abstract}
In this article, we take the point of view that the light scalar
meson $a_0(980)$ is a conventional $q\bar{q}$ state, and calculate
the coupling constants $g_{a_0\eta\pi^0}$ and $g_{a_0 \eta'\pi^0}$
with the light-cone QCD sum rules. The central value of the coupling
constant $g_{a_0\eta\pi^0}$ is consistent with the one extracted
from the radiative decay $\phi(1020)\rightarrow a_0(980)\gamma
\rightarrow \eta \pi^0 \gamma$.  The central value and lower bound
of the decay width $\Gamma_{a_{0}\rightarrow \eta
\pi^0}=127^{+84}_{-48}\, \rm{MeV} $ are compatible with the
experimental data of the total decay width
$\Gamma_{a_0(980)}=(50-100)\,\rm{MeV}$ from the Particle Data Group
with very model dependent estimation (the decay width can be much
larger), while the upper bound is too large. We give  possible
explanation for the discrepancy between the theoretical calculation
and experimental data.

\end{abstract}


\section{Introduction}
The light flavor  scalar mesons present a remarkable exception for
the constituent  quark models,  the structures of those mesons have
not been unambiguously determined yet
\cite{Godfray,Review1,Review2,Review3}. Experimentally, the strong
overlaps with each other
 and the broad widths (for the $f_0(980)$, $a_0(980)$, $f_0(1710)$, the widths are relatively  narrow)
 make their spectra cannot be approximated by the
  Breit-Wigner  formula.
  The numerous
candidates  with the same quantum numbers $J^{PC}=0^{++}$ below $2
\,\rm{GeV}$ can not be accommodated in one $q\bar{q}$ nonet,  some
are supposed to be glueballs, molecules and multiquark states
\cite{Review1,Review2,Review3}. The more elusive things are the
constituent structures of the mesons $f_0(980)$ and $a_0(980)$ with
almost the degenerate masses.

In the naive  quark model,
$a_0=(u\overline{u}-d\overline{d})/\sqrt{2}$ and
$f_0=s\overline{s}$; while in the framework of  the tetraquark
  models,  the  mesons $f_0(980)$ and $a_0(980)$
could either  be compact  objects (i.e. nucleon-like bound states of
quarks  with the symbolic quark structures  $f_0={s{\overline s}({ u
{\overline u}+d {\overline d})/ \sqrt{2}}}$ and $a_0=s {\overline
s}( u {\overline u}-d {\overline d}) / \sqrt{2}$
\cite{4quark1,4quark2}) or spatially  extended objects (i.e.
deuteron-like bound states of hadrons: $K {\overline K}$ molecules
\cite{KK1,KK2}). The hadronic dressing mechanism takes the point of
view that the mesons $f_0(980)$ and $a_0(980)$ have small $q\bar{q}$
cores of typical $q\bar{q}$ meson size,  strong couplings to the
intermediate hadronic states ($K \bar{K}$) enrich the pure
$q\bar{q}$ states with other components and spend part (or most
part) of their lifetime as virtual $ K \bar{K} $ states
\cite{HDress1,HDress2,HDress3}. In the hybrid model, those mesons
are tetraquark states $(qq)_{\bar{3}}(\bar{q}\bar{q})_3$  in the
$S$-wave near the center, with some constituents   $q \bar{ q}$ in
the $P$-wave, but further out they rearrange into  $(q \bar{ q})_1(q
\bar{ q})_1$ states and finally as meson-meson states
\cite{Review1,Review3}.
  All those interpretations have both outstanding
advantages and obvious shortcomings in one or other ways.

We can study the scalar mesons through  their couplings to two
pseudoscalar mesons,  two-photon decays and  radiative decays. The
radiative decays $\phi(1020)\rightarrow \pi^0 \pi^0\gamma$ and
$\phi(1020)\rightarrow \eta \pi^0 \gamma$
  have been the subject of intense investigation
  \cite{Radiative1,Radiative2,Radiative3,Radiative4,Radiative5,Radiative6,Radiative7}.
From the invariant $\pi^0 \pi^0$ and $\eta \pi^0$ mass
distributions, we can  obtain many  information about the nature of
the $f_0(980)$ and $a_0(980)$ respectively.

 In this article, we take the scalar mesons
$a_0(980)$ and $f_0(980)$ as the conventional $q\bar{q} $ states,
and calculate the values of the coupling constants
$g_{a_0\eta\pi^0}$ and $g_{a_0\eta'\pi^0}$  with the light-cone QCD
sum rules. The  coupling constant $g_{a_0 \eta \pi^0}$ is a basic
parameter in studying the radiative decay $\phi(1020)\rightarrow
a_0(980)\gamma\rightarrow\eta \pi^0 \gamma$. In previous works,  the
mesons $f_0(980)$, $a_0(980)$, $D_{s0}$, $D_{s1}$, $B_{s0}$ and
$B_{s1}$ were taken as the conventional $q\bar{q}$, $c\bar{s}$ and
$b\bar{s}$ states respectively, and the values of the coupling
constants $g_{f_0KK}$, $g_{a_0KK}$, $g_{D_{s0} DK}$, $g_{D_{s1}
D^*K}$, $g_{B_{s0} BK}$ and $g_{B_{s1} B^*K}$ have been  calculated
with the light-cone QCD sum rules
\cite{Wang23,Wang24,Wang06,Colangelo03,Wang04,Wang08}. The large
values of the coupling constants support the hadronic dressing
mechanism. In Ref.\cite{eta-Yilmaz}, the authors study the coupling
constant  $g_{a_0\eta\pi^0}$ with the interpolating current
$J^8_\mu=\frac{1}{\sqrt{6}}\left[\bar{u}\gamma_\mu \gamma_5
u+\bar{d}\gamma_\mu \gamma_5 d-2\bar{s}\gamma_\mu \gamma_5
s\right]$,  a complex subtraction procedure  is taken due to the
asymmetric  Borel parameters $M_1^2\neq M_2^2$. In this article, we
study the coupling constants  $g_{a_0\eta\pi^0}$ and
$g_{a_0\eta'\pi^0}$ together, and a simple subtraction procedure is
taken. The decay $f_0(980)\rightarrow \pi \pi$ can't occur at  the
tree level if the scalar meson $f_0(980)$ is a pure $s\bar{s} $
state, it should have some $n\bar{n} $ components,  the  coupling
constant $g_{f_0\pi\pi}$ has also been calculated with the
light-cone QCD sum rules \cite{f0-sigma}.

The light-cone QCD sum rules approach carries out the operator
product expansion near the light-cone $x^2\approx 0$ instead of the
short distance $x\approx 0$ while the nonperturbative matrix
elements are parameterized by the light-cone distribution amplitudes
  instead of
 the vacuum condensates \cite{LCSR1,LCSR2,LCSR3,LCSR4,LCSR5}. The nonperturbative
 parameters in the light-cone distribution amplitudes are calculated by
 the conventional QCD  sum rules
 and the  values are universal \cite{SVZ791,SVZ792,Reinders85}.

The article is arranged as follows: in section 2, we obtain the
coupling constants  $g_{a_0\eta\pi^0}$ and $g_{a_0\eta'\pi^0}$ with
the light-cone QCD sum rules; in section 3, numerical results;
section 4 is reserved for conclusion.

\section{ Coupling constants  $g_{a_0\eta\pi^0}$ and  $g_{a_0\eta'\pi^0}$ with light-cone QCD sum rules}
In the following, we write down the definitions   for the  coupling
constants  $g_{a_0\eta\pi^0}$ and  $g_{a_0\eta'\pi^0}$,
\begin{eqnarray}
\langle a_{0}|\eta
\pi^0\rangle&=&ig_{a_0\eta\pi^0}=i\sqrt{\frac{2}{3}}g \, \, , \nonumber\\
\langle a_{0}|\eta'
\pi^0\rangle&=&ig_{a_0\eta'\pi^0}=i\frac{2}{\sqrt{3}}g \, \, ,
\end{eqnarray}
where we have used the  phenomenological lagrangian $\mathcal {L}=g
\mbox{Tr}\left[ SPP\right]$, the $S$ and $P$ stand for the light
nonet scalar mesons and pseudoscalar mesons respectively. We study
the coupling constants $g_{a_0\eta\pi^0}$ and $g_{a_0\eta'\pi^0}$
with the
 two-point correlation function $\Pi_{\mu}(p,q)$,
\begin{eqnarray}
\Pi_{\mu}(p,q)&=&i\int d^4x e^{-iq\cdot x}\langle 0 |T\left\{J_{\mu}(0) J(x)\right\}|\pi^0(p)\rangle \, , \\
J_{\mu}(x)&=&{\bar u}(x)\gamma_\mu \gamma_5 u(x)+{\bar d}(x)\gamma_\mu \gamma_5 d(x)\, ,\nonumber  \\
J(x)&=&\frac{{\bar u}(x) u(x)-{\bar d}(x) d(x)}{\sqrt{2}}\, ,
\end{eqnarray}
where the currents $J_\mu(x)$ and $J(x)$ interpolate the
pseudoscalar mesons $\eta$, $\eta'$ and scalar meson $a_0(980)$,
respectively, the external $\pi^0$ meson has four momentum $p_\mu$
with $p^2=m_\pi^2$. One may think that it is more convenient to take
the octet current $J^8_{\mu}(x)$ and singlet current $J^0_{\mu}(x)$
\begin{eqnarray}
J^8_{\mu}(x)&=&\frac{{\bar u}(x)\gamma_\mu \gamma_5 u(x)+{\bar
d}(x)\gamma_\mu \gamma_5 d(x)-2{\bar s}(x)\gamma_\mu \gamma_5
s(x)}{\sqrt{6}} \, ,\nonumber\\
J^0_{\mu}(x)&=&\frac{{\bar u}(x)\gamma_\mu \gamma_5 u(x)+{\bar
d}(x)\gamma_\mu \gamma_5 d(x)+{\bar s}(x)\gamma_\mu \gamma_5
s(x)}{\sqrt{3}}
\end{eqnarray}
to interpolate the  pseudoscalar mesons $\eta$ and $\eta'$
respectively. The $\bar{s}s$ components of the interpolating
currents have no contributions at the level of quark-gluon degree's
of freedom, the octet current $J^8_{\mu}(x)$  and singlet current
$J^0_{\mu}(x)$ lead to the same analytical expressions. The
$J_\mu(x)$ is a linear composition  of  the  octet current
$J^8_{\mu}(x)$  and singlet current $J^0_{\mu}(x)$, we choose
 it to interpolate  the mesons  $\eta$ and $\eta'$ together,
\begin{eqnarray}
J_\mu(x)=\sqrt{\frac{2}{3}}J^8_\mu(x)+\frac{2}{\sqrt{3}}J^0_\mu(x)\,.
\end{eqnarray}
Despite which  interpolating currents  one may choose, the couplings
with the $a_0(980) \pi^0$  take place  through the $u\bar{u}$ and
$d\bar{d}$ components of the pseudoscalar mesons $\eta$ and $\eta'$
(not the $s\bar{s}$ component) at the level of quark-gluon degree's
of freedom. Although the  coupling constant $g_{a_0\eta'\pi^0}$ has
no direct  phenomenological interest, we take into account the
$\eta'$ meson to facilitate subtractions of the continuum states and
obtain more reliable QCD sum rules, we will revisit this subject at
the end of this section.

The correlation function $\Pi_{\mu}(p,q)$ can be decomposed as
\begin{eqnarray}
\Pi_{\mu}(p,q)&=&i \Pi(p,q) q_{\mu}+i\Pi_{A}(p,q)p_\mu
\end{eqnarray}
due to  Lorentz covariance, we choose the tensor structure
 $q_\mu$ for analysis.

According to the basic assumption of quark-hadron duality in the QCD
sum rules \cite{SVZ791,SVZ792,Reinders85}, we can insert  a complete
sets of intermediate hadronic states with the same quantum numbers
as the current operators $J_\mu(x)$  and $J(x)$ into the correlation
function $\Pi_{\mu}(p,q)$ to obtain the hadronic representation.
After isolating the ground state contributions from the pole terms
of the mesons $\eta$, $\eta'$ and $a_{0}(980)$, we get the following
result (we present some technical details in the appendix),
\begin{eqnarray}
  \Pi_{\mu}&=&\frac{\langle0| J_{\mu}(0)\mid \eta(q+p)\rangle }
  {M_{\eta}^2-(q+p)^2-i\epsilon}\langle
\eta|  J(0)| \pi^0\rangle+
 \frac{\langle0| J_{\mu}(0)\mid \eta'(q+p)\rangle}
  {M_{\eta'}^2-(q+p)^2-i\epsilon} \langle
\eta'|    J(0)|\pi^0\rangle + \cdots \nonumber \\
  &=&\frac{i2f_\eta(q+p)_\mu }
  {\sqrt{6}\left[M_{\eta}^2-(q+p)^2-i\epsilon\right]}\langle
\eta| a_0 \pi^0\rangle \frac{i}
  {q^2-M_{a_0}^2+i\epsilon} \langle a_0(q)|J(0)| 0\rangle+  \nonumber \\
  &&\frac{i2f_{\eta'}(q+p)_\mu }
  {\sqrt{3}\left[M_{\eta'}^2-(q+p)^2-i\epsilon\right]} \langle
\eta'| a_0 \pi^0\rangle\frac{ i  }
  {q^2-M_{a_0}^2+i\epsilon}\langle a_0(q)|J(0)| 0\rangle + \cdots \nonumber \\
  &=&\left[\frac{i 2g f_{\eta} f_{a_0}  M_{a_0}}
  {3\left[M_{\eta}^2-(q+p)^2-i\epsilon\right]\left[M_{a_0}^2-q^2-i\epsilon\right]}+\right.\nonumber\\
  &&\left.\frac{i 4g f_{\eta'} f_{a_0}  M_{a_0}}
  {3\left[M_{\eta'}^2-(q+p)^2-i\epsilon\right]\left[M_{a_0}^2-q^2-i\epsilon\right]}\right](p+q)_{\mu}
  + \cdots  ,
\end{eqnarray}
where the following definitions for the weak decay constants have
been used,
\begin{eqnarray}
\langle0 | J_{\mu}(0)|\eta(p)\rangle&=&\frac{i2f_{\eta}}{\sqrt{6}}p_\mu\,, \nonumber\\
\langle0 | J_{\mu}(0)|\eta'(p)\rangle&=&\frac{i2f_{\eta'}}{\sqrt{3}}p_\mu\,, \nonumber\\
\langle0 | J(0)|a_{0}(p)\rangle&=&f_{a_0}M_{a_0} \, .
\end{eqnarray}
We have take the ideal mixing limit for the $\eta$ and $\eta'$(i.e.
$\eta=|\frac{u\bar{u}+d\bar{d}-2s\bar{s}}{\sqrt{6}}\rangle$,
$\eta'=|\frac{u\bar{u}+d\bar{d}+s\bar{s}}{\sqrt{3}}\rangle$), and
neglect the anomaly contribution.

In the following, we briefly outline the  operator product expansion
for the correlation function  $\Pi_{\mu }(p,q)$ in perturbative QCD
theory. The calculations are performed at the large space-like
momentum regions $(q+p)^2\ll 0$ and  $q^2\ll 0$, which correspond to
the small light-cone distance $x^2\approx 0$ required by the
validity of the operator product expansion approach. We write down
the propagator of a massive quark in the external gluon field in the
Fock-Schwinger gauge firstly \cite{Belyaev94},
\begin{eqnarray}
S_{ij}(x_1,x_2) &=&
 i \int\frac{d^4k}{(2\pi)^4}e^{-ik(x_1-x_2)}\nonumber\\
 && \left\{
\frac{\not\!k +m}{k^2-m^2} \delta_{ij} -\int\limits_0^1 dv\, g_s \,
G^{\mu\nu}_{ij}(vx_1+(1-v)x_2) \right. \nonumber \\
&&\left. \Big[ \frac12 \frac {\not\!k
+m}{(k^2-m^2)^2}\sigma_{\mu\nu} - \frac1{k^2-m^2}v(x_1-x_2)_\mu
\gamma_\nu \Big]\right\}\, .
\end{eqnarray}
 Substituting the  $u$ and $d$ quark propagators and the corresponding $\pi$-meson
light-cone distribution amplitudes into the correlation function
$\Pi_{\mu}(p,q)$,  and completing the integrals over the variables
$x$ and $k$, finally we obtain the analytical expressions. In
calculation, the two-particle and three-particle $\pi$-meson
light-cone distribution amplitudes have been used
\cite{PSLC1,PSLC2,PSLC3,PSLC4}, the explicit expressions are given
in the appendix. The parameters in the light-cone distribution
amplitudes are scale dependent and are estimated with the QCD sum
rules \cite{PSLC1,PSLC2,PSLC3,PSLC4}. In this article, the energy
scale $\mu$ is chosen to be $\mu=1\,\rm{GeV}$.

After straightforward calculations, we obtain the final expression
of the double Borel transformed correlation function $\Pi$ at the
level of quark-gluon degrees of freedom. The masses of the
pseudoscalar meson and scalar meson are $M_{\eta'} =0.958\,\rm{GeV}$
and $M_{a_0} =0.985\,\rm{GeV}$ respectively,
\begin{eqnarray}
 \frac{M_{\eta'}^2}{M_{\eta'}^2+M_{a_0}^2}\approx0.49 \, ,
\end{eqnarray}
 there exists an overlapping working window for the two Borel
parameters $M_1^2$ and $M_2^2$, it is convenient to take the value
$M_1^2=M_2^2$, $M^2=\frac{M_1^2M_2^2}{M_1^2+M_2^2}$. We introduce
the threshold parameter $s_0$ and make the simple replacement,
\begin{eqnarray}
e^{-\frac{m_u^2+u_0(1-u_0)m_\pi^2}{M^2}} \rightarrow
e^{-\frac{m_u^2+u_0(1-u_0)m_\pi^2}{M^2} }-e^{-\frac{s_0}{M^2}}
\nonumber
\end{eqnarray}
 to subtract the contributions from the high resonances  and
  continuum states \cite{Belyaev94}. Finally we obtain the sum rule for the  coupling
  constant $g$,
\begin{eqnarray}
g&=& \frac{3\exp\left( \frac{M^2_{a_0}}{M_2^2}
\right)}{2f_{a_0}M_{a_0}\left[ f_\eta\exp\left(-
\frac{M^2_{\eta}}{M_1^2} \right)+2f_\eta'\exp\left(-
\frac{M^2_{\eta'}}{M_1^2} \right) \right]}\left\{\left[\exp\left(-
\frac{\Xi}{M^2}\right)-\exp\left(-
\frac{s_0}{M^2}\right)\right]  \right.\nonumber\\
&& \frac{f_\pi
m_\pi^2M^2}{2m_u}\left[\varphi_p(u_0)-\frac{d\varphi_\sigma(u_0)}{6du_0}\right]
+\exp\left(-\frac{\Xi}{M^2}\right)\left[ -m_uf_\pi m_\pi^2
\int_0^{u_0}
dt B(t)  \right.\nonumber\\
  &&+f_{3\pi}m_\pi^2 \int_0^{u_0}d\alpha_u
  \int_{u_0-\alpha_u}^{1-\alpha_u}d\alpha_g
\varphi_{3\pi}(1-\alpha_u-\alpha_g,\alpha_g,\alpha_u)\frac{2(\alpha_u+\alpha_g-u_0)-3\alpha_g
}{\alpha_g^2}
\nonumber\\
&&-\frac{2m_uf_\pi m_\pi^4}{M^2}  \int_{1-u_0}^1 d\alpha_g
\frac{1-u_0}{\alpha_g^2}\int_0^{\alpha_g}
d\beta\int_0^{1-\beta}d\alpha \Phi(1-\alpha-\beta,\beta,\alpha)
\nonumber \\
&& +\frac{2m_uf_\pi m_\pi^4}{M^2}\left(\int_0^{1-u_0} d\alpha_g
\int^{u_0}_{u_0-\alpha_g} d\alpha_u \int_0^{\alpha_u} d\alpha
+\int^1_{1-u_0} d\alpha_g \int^{1-\alpha_g}_{u_0-\alpha_g} d\alpha_u
\int_0^{\alpha_u} d\alpha\right) \nonumber\\
&&\left.\left.\frac{\Phi(1-\alpha-\alpha_g,\alpha_g,\alpha)}{\alpha_g}
\right]\right\} \, ,
\end{eqnarray}
where
\begin{eqnarray}
\Phi(\alpha_i)&=&A_\parallel(\alpha_i)+A_\perp(\alpha_i)-V_\parallel(\alpha_i)-V_\perp(\alpha_i) \, ,\nonumber \\
\Xi&=&m_u^2+u_0(1-u_0)m_\pi^2 \, ,\nonumber \\
u_0&=&\frac{M_1^2}{M_1^2+M_2^2}\, ,
\end{eqnarray}
and we have taken the isospin limit $m_u=m_d$.

In Ref.\cite{eta-Yilmaz} (also in
Refs.\cite{Colangelo03,Wang04,f0-sigma}),  a complex subtraction
procedure  is taken  due to the asymmetry Borel parameters,
$M_1^2\neq M_2^2$. In the light-cone QCD sum rules, we often take
the technique developed in Ref.\cite{Belyaev94} to obtain the
spectral densities at the level of quark-gluon degrees of freedom,
\begin{eqnarray}
 \Pi&=& \int_0^1 \frac{ f(u)}{\Delta-(q+up)^2}du \, \nonumber\\
 &=&\int_{\Delta}^{\infty} \frac{\rho_{\rm{QCD}}(s)}{\left[s-(p+q)^2\right]\left[s-q^2\right]}ds \nonumber\\
 &=&\int_{\Delta_1}^{\infty}\int_{\Delta_2}^{\infty}
 \frac{\rho_{\rm{QCD}}(s_1,s_2)\delta(s_1-s_2)}{\left[s_1-(p+q)^2\right]\left[s_2-q^2\right]}ds_1ds_2
 \, ,
\end{eqnarray}
where the $f(u)$ are functions of the two-particle light-cone
distribution amplitudes, $u=\frac{\Delta-q^2}{s-q^2}$, the $\Delta$
are the squared masses of the exchanged quarks, the $\Delta_1$ and
$\Delta_2$ are the corresponding thresholds. It works
 efficiently   in the case where the threshold parameters
$s_1^0$ and $s^0_2$  differ from each other slightly.  If we take
the values $s_1^0=s_\eta^0=(0.7-0.8)\,\rm{GeV}^2$ (in the case that
the octet current $J_\mu^8(x)$ is chosen to interpolate the $\eta$
meson, see Ref.\cite{eta-Yilmaz}) and
$s_2^0=s_{a_0}^0>M_{a_0}^2\approx 1\, \rm{GeV}^2$, the contributions
from the $a_0(980)$ are not  taken into account properly,
\begin{eqnarray}
 \Pi&=&\int_{\Delta_1}^{s^0_1}\int_{\Delta_2}^{s_2^0} \frac{\rho_{\rm{QCD}}(s_1,s_2)\delta(s_1-s_2)}{\left[s_1-(p+q)^2\right]\left[s_2-q^2\right]}ds_1ds_2
 +\cdots \nonumber \\
  &=&\int_{\Delta_1}^{s^0_1}\int_{\Delta_2}^{s_1^0} \frac{\rho_{\rm{QCD}}(s_1,s_2)\delta(s_1-s_2)}{\left[s_1-(p+q)^2\right]\left[s_2-q^2\right]}ds_1ds_2
 +\cdots\, .
 \end{eqnarray}

In the case of non-equal threshold parameters $s_1^0\neq s_2^0$, we
can take
 $s_0=\rm{max}(s_1^0,s^0_2)$ with $s_0$ small enough to avoid the contaminations from
 the high resonances in either of the two channels,   or take $s_0=\rm{min}(s_1^0,s^0_2)$ with $s_0$ large enough to
 take into account the contributions  from the ground states in either of the two
 channels. We have two choices in general, which can result in some
 uncertainties.  In this article, we choose the
current $J_\mu(x)$ to interpolate both the $\eta$ and $\eta'$ mesons
to overcome the shortcoming, and take into account  the
contributions from the $\eta'$ meson at the phenomenological side.

\section{Numerical result and discussion}
The input parameters of the light-cone distribution amplitudes are
taken as  $\lambda_3=0.0$, $f_{3\pi}=(0.45\pm0.15)\times
10^{-2}\,\rm{GeV}^2$, $\omega_3=-1.5\pm0.7$, $\omega_4=0.2\pm0.1$,
$a_1=0.0 $, $a_2=0.28\pm 0.08$,  $a_4=0.0 $,  $\eta_4=10.0\pm3.0$
\cite{PSLC1,PSLC2,PSLC3,PSLC4}, $m_u=m_d=m_q=(5.6\pm 1.6)
\,\rm{MeV}$, $f_\pi=0.130\,\rm{GeV}$, $m_{\pi} =0.135\,\rm{GeV}$,
$M_{\eta} =0.547\,\rm{GeV}$, $M_{\eta'} =0.958\,\rm{GeV}$, $M_{a_0}
=0.985\,\rm{GeV}$,  $f_{\eta}=1.3f_{\pi}$, $f_{\eta'}=1.2f_{\pi}$
\cite{decay-eta}, and $f_{a_0}=(0.21\pm 0.01)\,\rm{GeV}$
\cite{Wang04}.

The axial-vector current $J_\mu(x)$ has also non-vanishing couplings
with both the pseudoscalar mesons $\eta(1295)$, $\eta(1405)$,
$\eta(1475)$, etc and the axial-vector mesons $f_1(1285)$, etc. The
scalar current $J(x)$ has also non-vanishing couplings with the
scalar mesons $a_0(1450)$, etc. The masses and widths of those
mesons are  $M_{\eta(1295)}=(1294\pm4)\,\rm{MeV}$,
$\Gamma_{\eta(1295)}=(55\pm 5)\,\rm{MeV}$;
$M_{\eta(1405)}=(1409.8\pm2.5)\,\rm{MeV}$,
$\Gamma_{\eta(1405)}=(51.1\pm 3.4)\,\rm{MeV}$;
$M_{\eta(1475)}=(1476\pm4)\,\rm{MeV}$, $\Gamma_{\eta(1475)}=(87\pm
9)\,\rm{MeV}$; $M_{f_1(1285)}=(1281.8\pm 0.6)\,\rm{MeV}$,
$\Gamma_{f_1(1285)}=(24.2\pm 1.1)\,\rm{MeV}$;
$M_{a_0(1450)}=(1474\pm19)\,\rm{MeV}$ and
$\Gamma_{a_0(1450)}=(265\pm13)\,\rm{MeV}$ from the Particle Data
Group \cite{PDG}.

From the experimental data, we can see that the $a_0$ channel
permits larger threshold parameter  than that of the $\eta$ channel.
If we take the value $s_0=\rm{max}
(s^0_\eta,s^0_{a_0})=s_{a_0}^0\leq 1.7 \, \rm{GeV}^2$, the
contaminations from the $\eta(1295)$ and $f_1(1285)$ are included
in. We have to take the other choice, $s_0=\rm{min}
(s^0_\eta,s^0_{a_0})= s^0_\eta\leq1.6 \, \rm{GeV}^2$.  It happens to
be the ideal choice and reproduces  the mass of the $a_0(980)$ with
the conventional two-point QCD sum rules for the Borel parameter 
$M^2=(1.0-1.6)\, \rm{GeV}^2$.

In this article, we take the threshold parameter and Borel parameter
as  $s_0=(1.4-1.6)\,\rm{GeV}^2$ and $M^2=(1.0-1.6)\, \rm{GeV}^2$ to
avoid the contaminations from the high resonances and continuum
states as $\exp\left(-\frac{s_0}{M^2}\right)=0.2-0.4$. In this
region, the value of the coupling constant $g$ is rather stable with
variation of the Borel parameter, see Figs.(1-2).

In this article, we take the values of the coefficients $a_i$ of the
twist-2 light-cone distribution amplitude $\varphi_\pi(u)$ from the
conventional QCD sum rules \cite{PSLC1,PSLC4}. The $\varphi_\pi(u)$
has been  analyzed with the light-cone QCD sum rules and (non-local
condensates) QCD sum rules confronting with the high precision CLEO
data on the $  \gamma \gamma^* \to \pi^0$ transition form-factor
\cite{Schmedding1999-24,Bakulev2001-24,Bakulev2002-24,Bakulev2003-24,Bakulev2006-24,Bakulev2006-24-2}.
We also study the coupling constants $g_{a_0\eta\pi^0}$ and
$g_{a_0\eta'\pi^0}$ with the values $a_2=0.29$ and $a_4=-0.21$ at
$\mu=1$ GeV, which are obtained via one-loop renormalization group
equation for the central values $a_2=0.268$ and $a_4=-0.186$ at
$\mu^2=1.35\,\,\rm{GeV}^2$ from the (non-local condensates) QCD sum
rules with improved model \cite{Bakulev2006-24-2}.

 In the limit of large Borel parameter $M^2$, the  coupling
constant $g$ takes up the following behavior,
\begin{eqnarray}
g&\propto& \frac{ M^2}{m_u} \left[
\varphi_p(u_0)-\frac{d\varphi_\sigma(u_0)}{6du_0}\right]\, .
\end{eqnarray}
It is not unexpected, the contributions  from the two-particle
twist-3 light-cone distribution amplitude $\varphi_p(u)$ are greatly
enhanced by the large Borel parameter $M^2$,  (large) uncertainties
of the relevant parameters presented in above equations have
significant impact on the numerical results. The contribution from
the two-particle  twist-3  $\varphi_\sigma(u_0)$ is zero due to
symmetry property. If we take the value $m_u=m_d=m_q=(5.6\pm 1.6)
\,\rm{MeV}$ \cite{PSLC4}, the uncertainty comes from the $m_q$ is
very large, about $(33-64)\%$, and the predictive ability is poor,
see Fig.1.
\begin{figure}
\centering
  \includegraphics[totalheight=8cm,width=10cm]{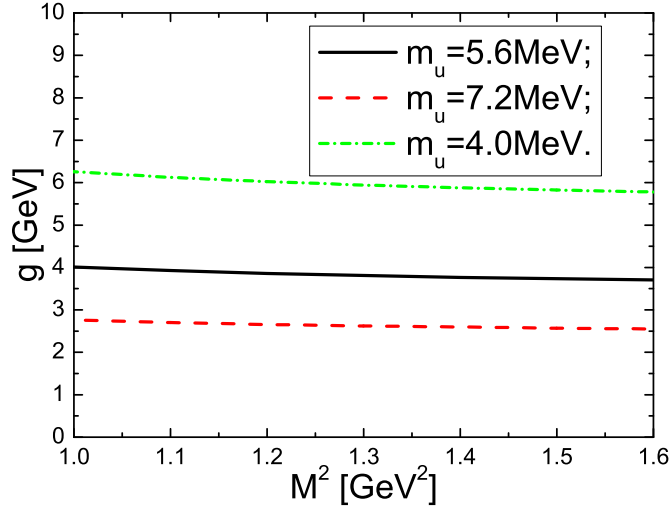}
      \caption{The  coupling constant  $g$  with the parameters $M^2$ and $m_u$. }
\end{figure}
From the Gell-Mann-Oakes-Renner relation, we can obtain
$\frac{f_\pi^2m_\pi^2}{m_u+m_d}=(0.027\pm0.003)\,\rm{GeV}^3$
\cite{PSLC1}, i.e. $m_q\approx(5.6\pm0.6)\,\rm{MeV}$, which can
result in much smaller uncertainty.

Taking into account all the uncertainties of the input parameters,
finally we obtain the numerical values of the  coupling constants,
which are shown in Fig.2,
\begin{eqnarray}
  g &=&3.8_{-1.4}^{+2.5} \,\rm{GeV} \,  , \nonumber\\
  g_{a_0 \eta \pi^0} &=&3.1_{-1.1}^{+2.0} \,\rm{GeV} \, ,\nonumber \\
g_{a_0 \eta' \pi^0} &=&4.4_{-1.6}^{+2.9} \,\rm{GeV}  \, ,
 \end{eqnarray}
for $ m_q =(5.6\pm 1.6) \,\rm{MeV}$ and
\begin{eqnarray}
  g &=&3.8_{-0.8}^{+1.1} \,\rm{GeV} \, , \nonumber\\
  g_{a_0 \eta \pi^0} &=&3.1_{-0.7}^{+0.9} \,\rm{GeV} \, ,\nonumber \\
g_{a_0 \eta' \pi^0} &=&4.4_{-0.9}^{+1.3} \,\rm{GeV}  \, ,
 \end{eqnarray}
 for $ m_q =(5.6\pm 0.6) \,\rm{MeV} $.
The  parameters of the twist-2 light-cone distribution amplitude
 $\varphi_\pi(u)$ obtained in Ref.\cite{Bakulev2006-24-2} can change
  the value of the  coupling constant
slightly, less than  $0.1\%$.

In table 1, we list  the values (not all) of the  coupling constant
$g_{a_0\eta\pi^0}$ from different quark models and the experimental
data. From the table, we can see that the values of the early
estimations  with the $q\bar{q}$ model, tetraquark  model and
$K\bar{K}$ molecule model deviate greatly from the experimental data
\cite{KLOE1,KLOE2,KLOE07}, we can't use them to identify the
structures of the $a_0(980)$ with confidence.
 Comparing with the values   extracted from
 the radiative decay $\phi(1020)\rightarrow a_0(980)\gamma
\rightarrow \eta \pi^0 \gamma$ \cite{SND1,SND2,KLOE1,KLOE2,KLOE07},
the central value of our numerical result is reasonable and support
the $q\bar{q}$ model.

\begin{table}[ht]
\centering
\begin{tabular}{|c|c|}
\hline  quark models and experimental data & $g_{a_{0}\eta\pi^0}(\rm{GeV})$ \\
\hline
  $q\bar{q}$ model \cite{Achasov89}  & $2.03$ \\
\hline
  tetraquark model \cite{Achasov89}  & $4.57$ \\
\hline
 $K\bar{K}$ molecule model \cite{Achasov97,KK2}  & $1.74$ \\
\hline
 SND Collaboration \cite{SND1,SND2} & $3.11$ \\
\hline
 KLOE Collaboration \cite{KLOE1,KLOE2}  & $3.0\pm0.2$ \\
\hline
 KLOE Collaboration \cite{KLOE07}  & $2.8\pm0.1$ \\
\hline
light-cone sum rules ($q\bar{q}$ model)\cite{eta-Yilmaz}  & $2.6-3.4$ \\
\hline
This work ($q\bar{q}$ model) & $3.1_{-0.7}^{+0.9}$ \\
\hline
\end{tabular}
\caption{  The  coupling constant $g_{a_{0}\eta \pi^0}$ from
different quark models and experimental data.}
\end{table}

 From the coupling constant
$g_{a_0\eta\pi^0}$, we can obtain the decay width
$\Gamma_{a_{0}\rightarrow \eta \pi^0}$,
\begin{eqnarray}
   \Gamma_{a_{0}\rightarrow \eta \pi^0} &=&\frac{pg_{a_0\eta\pi^0}^2}{8\pi M_{a_0}^2} \, , \\
     &=&127^{+222}_{-76} \,\rm{MeV} \,\, \rm{for}\,\, g =3.8_{-1.4}^{+2.5} \,\rm{GeV}  \, , \nonumber \\
     &=&127^{+84}_{-48} \,\rm{MeV} \,\, \rm{for}\,\, g =3.8_{-0.8}^{+1.1}
     \,\rm{GeV}\, ,\nonumber\\
    p&=&\frac{\sqrt{\left[M_{a_0}^2-(M_{\eta}+m_\pi)^2\right]\left[M_{a_0}^2-(M_{\eta}-m_\pi)^2\right]}}{2M_{a_0}}
  \, . \nonumber
 \end{eqnarray}
Comparing  with the experimental data
$\Gamma_{a_0(980)}=(50-100)\,\rm{MeV}$ from the Particle Data Group
with  very model dependent estimation (the decay width can be much
larger) \cite{PDG}, the central value and lower bound of our
numerical result $\Gamma_{a_{0}\rightarrow \eta
\pi^0}=127^{+84}_{-48} \,\rm{MeV} $ are reasonable, however,  the
upper-bound  is too large, we should reduce the uncertainties of the
input parameters $f_{3\pi}$ and $m_q$ (main uncertainties originate
from  them) before make definite conclusion.

In this article, we take the point of view that the $a_0(980)$ is a
scalar $q\bar{q} $ state. In Ref.\cite{4Coupling}, the light nonet
scalar mesons are taken as tetraquark states, and the coupling
constants among the light scalar mesons and pseudoscalar mesons are
calculated with the QCD sum rules. The numerical results indicate
that the values of the coupling constants for the tetraquark states
are always smaller  than the corresponding ones for the $q\bar{q} $
states \cite{Colangelo03,Wang04}.

The predictions listed in Table 1 are obtained from the
phenomenological (potential) quark models
\cite{KK2,Achasov89,Achasov97}, and the resulting  coupling constant
$g$ differs  from the  corresponding ones from the QCD sum rules
greatly \cite{Colangelo03,Wang04,4Coupling}. Furthermore, those
predictions also differ from the ones extracted from the
experimental data significantly \cite{SND1,SND2,KLOE1,KLOE2,KLOE07}.
In this article, we prefer the values from the QCD sum rules for
consistence, i.e. if the nonet scalar mesons are tetraquark states,
they have much smaller coupling constant $g$
\cite{Colangelo03,Wang04,4Coupling}.

The scalar meson $a_0(980)$ may have small $q\bar{q} $ kernel of the
typical $q\bar{q} $ meson size, strong coupling  to the nearby
$\bar{K}K$ threshold may result in some tetraquark components,
whether the nucleon-like bound state or deuteron-like bound state.
The
 tetraquark components  may lead to smaller decay width, and smear the discrepancy between the
(upper bound of) theoretical calculation and the experimental data.

\begin{figure} \centering
  \includegraphics[totalheight=6cm,width=7cm]{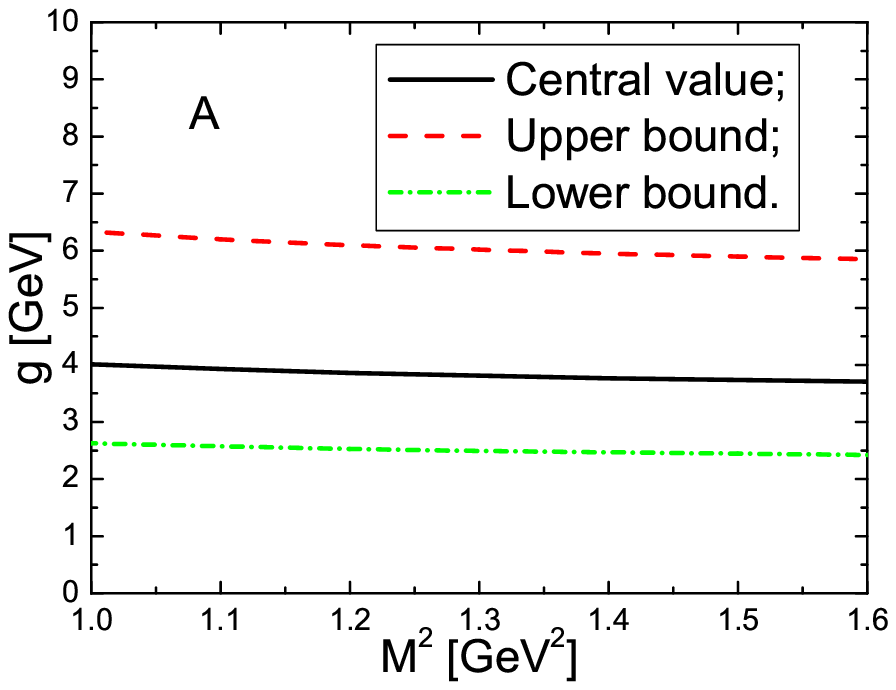}
 \includegraphics[totalheight=6cm,width=7cm]{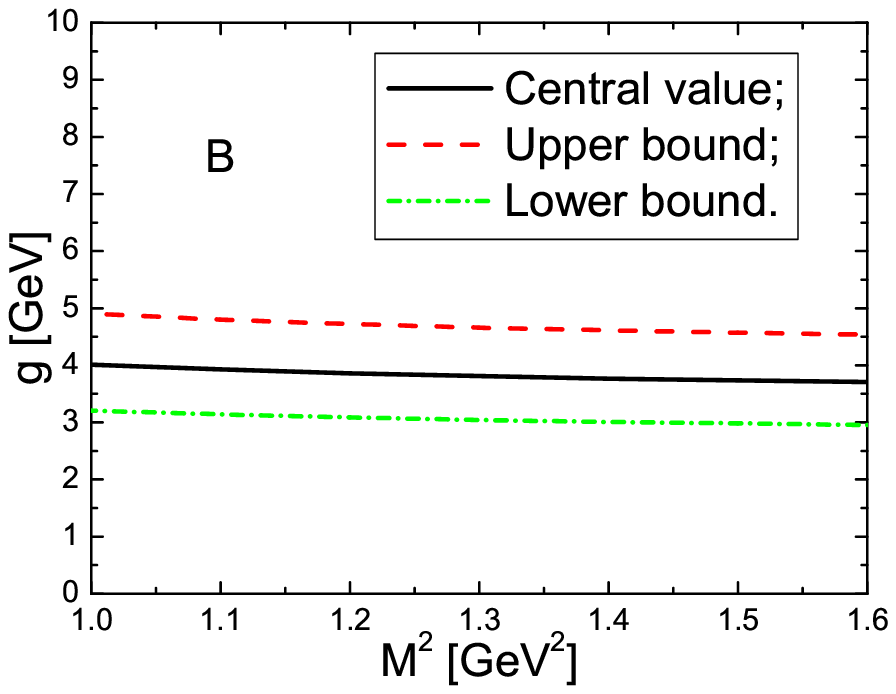}
     \caption{The  coupling constant $g$ with the parameter $M^2$,
     A for $m_q=(5.6\pm1.6) \rm{MeV}$ and B for $m_q=(5.6\pm0.6) \,\rm{MeV}$. }
\end{figure}

\section{Conclusion}

In this article, we take the point of view that the scalar meson
$a_0(980)$ is a conventional $q\bar{q} $ state, and calculate the
coupling constants $g_{a_0\eta\pi^0}$ and $g_{a_0 \eta'\pi^0}$ with
the light-cone QCD sum rules.  Although the coupling constant
$g_{a_0\eta'\pi^0}$ has no direct phenomenological interest, we take
into account the $\eta'$ meson to facilitate subtraction of the
continuum states to give more reliable sum rule.
 The central value of  the coupling constant
$g_{a_0\eta\pi^0}$ is consistent with  the values  extracted from
the radiative decay $\phi(1020)\rightarrow a_0(980)\gamma
\rightarrow \eta \pi^0 \gamma$.   The central value and lower bound
of  the decay width $\Gamma_{a_{0}\rightarrow \eta
\pi^0}=127^{+84}_{-48} \,\rm{MeV} $ are compatible  with the
experimental data of the total decay width
$\Gamma_{a_0(980)}=(50-100)\,\rm{MeV}$ from  the Particle Data Group
with very model dependent estimation (the decay width can be much
larger), while the upper bound is too large. The scalar meson
$a_0(980)$ may have small $q\bar{q} $ kernel of the typical
$q\bar{q}$ meson size, strong coupling to the nearby $\bar{K}K$
threshold may result in some tetraquark components, whether the
nucleon-like bound state  or
 deuteron-like bound state.  The
 tetraquark components  may lead to smaller decay width, and smear the discrepancy between the
theoretical calculation and the experimental data.

\section*{Acknowledgment}
This  work is supported by National Natural Science Foundation,
Grant Number 10775051, and Program for New Century Excellent Talents
in University, Grant Number NCET-07-0282.

\section*{Appendix}
We  present some technical details in obtaining the spectral density
at the phenomenological side,
\begin{eqnarray}
\langle \eta|  J(0)| \pi^0\rangle &=&\langle \eta(p')|\sum_a\int
\frac{d^3\vec{q}}{(2\pi)^3 2E}|a(q) \rangle \langle a(q)| J(0)|
\pi^0(p)\rangle \nonumber\\
&=&\sum_a\int \frac{d^4q}{(2\pi)^4 }\langle
\eta(p')|\frac{i}{q^2-M_a^2+i\epsilon}|a(q) \rangle
 \langle a(q)| J(0)| \pi^0(p)\rangle\nonumber\\
&=&\sum_a\int \frac{d^4q}{(2\pi)^4 }\langle \eta(p')|a(q)
\pi^0(p)\rangle \frac{i}{q^2-M_a^2+i\epsilon} \langle a(q)| J(0)|0
\rangle\nonumber\\
&=&\sum_af_aM_a\int \frac{d^4 q}{(2\pi)^4 }\langle \eta(p')|i\int
d^4y \, \mathcal{L}(y) |a(q) \pi^0(p)\rangle
\frac{i}{q^2-M_a^2+i\epsilon}
 \nonumber\\
&=&\sum_a\int \frac{d^4 q}{(2\pi)^4 }\langle \eta(p')|\int d^4y\,
\mathbbm{g}_a \eta(y)a(y)\pi^0(y) |a(q) \pi^0(p)\rangle
\frac{f_aM_a}{M_a^2-q^2-i\epsilon}\nonumber\\
&=&\sum_a\int \frac{d^4 q}{(2\pi)^4 } (2\pi)^4\delta^4(p'-p-q)
\mathbbm{g}_a\frac{f_aM_a}{M_a^2-q^2-i\epsilon}\nonumber\\
&=& \sum_a\frac{\mathbbm{g}_af_aM_a}{M_a^2-(p'-p)^2-i\epsilon} \, ,
\nonumber
\end{eqnarray}
where we have used the completeness relation,
\begin{eqnarray}
\sum_a\int \frac{d^3\vec{q}}{(2\pi)^3 2E}|a(q) \rangle \langle
a(q)|=1 \, , \nonumber
\end{eqnarray}
which corresponds to the normalization condition $\langle
a(q)|a(q')\rangle=(2\pi)^3 2E\delta^3(\vec{q}-\vec{q'})$, the $a$'s
are the intermediate hadronic states with the same quantum numbers
as the current operator $J(0)$, the $\mathbbm{g}_a$ denote the
corresponding coupling constant among the $\eta$, $a$ and $\pi^0$,
and $\langle0 | J(0)|a(q)\rangle=f_{a}M_{a}$. In the light-cone QCD
sum rules, we often take the economical routine,
\begin{eqnarray}
\langle \eta|  J(0)| \pi^0\rangle&=&\sum_a\langle \eta(p')|a(q)
\pi^0(p)\rangle \frac{i}{q^2-M_a^2+i\epsilon} \langle a(q)| J(0)|0
\rangle\nonumber\\
&=& \sum_a\frac{\mathbbm{g}_af_aM_a}{M_a^2-q^2-i\epsilon} \, ,
\nonumber
\end{eqnarray}
with a suitable definition $\langle \eta(p')|a(q)
\pi^0(p)\rangle=i\mathbbm{g}_a$.

The light-cone distribution amplitudes of the $\pi$ meson are
defined as,
\begin{eqnarray}
\langle0| {\bar u} (0) \gamma_\mu \gamma_5 d(x) |\pi(p)\rangle& =& i
f_\pi p_\mu \int_0^1 du  e^{-i u p\cdot x}
\left\{\varphi_\pi(u)+\frac{m_\pi^2x^2}{16}
A(u)\right\}\nonumber\\
&&+if_\pi m_\pi^2\frac{x_\mu}{2p\cdot x}
\int_0^1 du  e^{-i u p \cdot x} B(u) \, , \nonumber\\
\langle0| {\bar u} (0) i \gamma_5 d(x) |\pi(p)\rangle &=&
\frac{f_\pi m_\pi^2}{ m_u+m_d}
\int_0^1 du  e^{-i u p \cdot x} \varphi_p(u)  \, ,  \nonumber\\
\langle0| {\bar u} (0) \sigma_{\mu \nu} \gamma_5 d(x) |\pi(p)\rangle
&=&i(p_\mu x_\nu-p_\nu x_\mu)  \frac{f_\pi m_\pi^2}{6 (m_u+m_d)}
\int_0^1 du
e^{-i u p \cdot x} \varphi_\sigma(u) \, ,  \nonumber\\
\langle0| {\bar u} (0) \sigma_{\alpha \beta} \gamma_5 g_s G_{\mu
\nu}(v x)d(x) |\pi(p)\rangle&=& f_{3 \pi}\left\{(p_\mu p_\alpha
g^\bot_{\nu
\beta}-p_\nu p_\alpha g^\bot_{\mu \beta}) -(p_\mu p_\beta g^\bot_{\nu \alpha}\right.\nonumber\\
&&\left.-p_\nu p_\beta g^\bot_{\mu \alpha})\right\} \int {\cal
D}\alpha_i \varphi_{3 \pi} (\alpha_i)
e^{-ip \cdot x(\alpha_d+v \alpha_g)} \, ,\nonumber\\
\langle0| {\bar u} (0) \gamma_{\mu} \gamma_5 g_s G_{\alpha
\beta}(vx)d(x) |\pi(p)\rangle&=&  p_\mu  \frac{p_\alpha
x_\beta-p_\beta x_\alpha}{p
\cdot x}f_\pi m_\pi^2\nonumber\\
&&\int{\cal D}\alpha_i A_{\parallel}(\alpha_i) e^{-ip\cdot
x(\alpha_d +v \alpha_g)}\nonumber \\
&&+ f_\pi m_\pi^2 (p_\beta g_{\alpha\mu}-p_\alpha
g_{\beta\mu})\nonumber\\
&&\int{\cal D}\alpha_i A_{\perp}(\alpha_i)
e^{-ip\cdot x(\alpha_d +v \alpha_g)} \, ,  \nonumber\\
\langle0| {\bar u} (0) \gamma_{\mu}  g_s \tilde G_{\alpha
\beta}(vx)d(x) |\pi(p)\rangle&=& p_\mu  \frac{p_\alpha
x_\beta-p_\beta x_\alpha}{p \cdot
x}f_\pi m_\pi^2\nonumber\\
&&\int{\cal D}\alpha_i V_{\parallel}(\alpha_i) e^{-ip\cdot
x(\alpha_d +v \alpha_g)}\nonumber \\
&&+ f_\pi m_\pi^2 (p_\beta g_{\alpha\mu}-p_\alpha
g_{\beta\mu})\nonumber\\
&&\int{\cal D}\alpha_i V_{\perp}(\alpha_i) e^{-ip\cdot x(\alpha_d +v
\alpha_g)} \, ,
\end{eqnarray}
where  $\tilde G_{\alpha \beta}= {1\over 2} \epsilon_{\alpha \beta
\mu\nu} G^{\mu\nu} $ and ${\cal{D}} \alpha_i =d \alpha_1 d \alpha_2
d \alpha_3 \delta(1-\alpha_1 -\alpha_2 -\alpha_3)$.

The  light-cone distribution amplitudes are parameterized as
\begin{eqnarray}
\varphi_\pi(u)&=&6u(1-u)
\left\{1+a_1C^{\frac{3}{2}}_1(2u-1)+a_2C^{\frac{3}{2}}_2(2u-1)+a_4C^{\frac{3}{2}}_4(2u-1)
\right\}\, , \nonumber\\
\varphi_p(u)&=&1+\left\{30\eta_3-\frac{5}{2}\rho^2\right\}C_2^{\frac{1}{2}}(2u-1)\nonumber \\
&&+\left\{-3\eta_3\omega_3-\frac{27}{20}\rho^2-\frac{81}{10}\rho^2 a_2\right\}C_4^{\frac{1}{2}}(2u-1)\, ,  \nonumber \\
\varphi_\sigma(u)&=&6u(1-u)\left\{1
+\left[5\eta_3-\frac{1}{2}\eta_3\omega_3-\frac{7}{20}\rho^2-\frac{3}{5}\rho^2 a_2\right]C_2^{\frac{3}{2}}(2u-1)\right\}\, , \nonumber \\
\varphi_{3\pi}(\alpha_i) &=& 360 \alpha_u \alpha_d \alpha_g^2 \left
\{1 +\lambda_3(\alpha_u-\alpha_d)+ \omega_3 \frac{1}{2} ( 7 \alpha_g
- 3) \right\} \, , \nonumber\\
V_{\parallel}(\alpha_i) &=& 120\alpha_u \alpha_d \alpha_g \left(
v_{00}+v_{10}(3\alpha_g-1)\right)\, ,
\nonumber \\
A_{\parallel}(\alpha_i) &=& 120 \alpha_u \alpha_d \alpha_g a_{10}
(\alpha_d-\alpha_u)\, ,
\nonumber\\
V_{\perp}(\alpha_i) &=& -30\alpha_g^2
\left\{h_{00}(1-\alpha_g)+h_{01}\left[\alpha_g(1-\alpha_g)-6\alpha_u
\alpha_d\right] \right.  \nonumber\\
&&\left. +h_{10}\left[
\alpha_g(1-\alpha_g)-\frac{3}{2}\left(\alpha_u^2+\alpha_d^2\right)\right]\right\}\,
, \nonumber\\
A_{\perp}(\alpha_i) &=&  30 \alpha_g^2 (\alpha_u-\alpha_d) \left\{h_{00}+h_{01}\alpha_g+\frac{1}{2}h_{10}(5\alpha_g-3)  \right\}, \nonumber\\
A(u)&=&6u(1-u)\left\{
\frac{16}{15}+\frac{24}{35}a_2+20\eta_3+\frac{20}{9}\eta_4 \right.
\nonumber \\
&&+\left[
-\frac{1}{15}+\frac{1}{16}-\frac{7}{27}\eta_3\omega_3-\frac{10}{27}\eta_4\right]C^{\frac{3}{2}}_2(2u-1)
\nonumber\\
&&\left.+\left[
-\frac{11}{210}a_2-\frac{4}{135}\eta_3\omega_3\right]C^{\frac{3}{2}}_4(2u-1)\right\}+\left\{
 -\frac{18}{5}a_2+21\eta_4\omega_4\right\} \nonumber\\
 && \left\{2u^3(10-15u+6u^2) \log u+2\bar{u}^3(10-15\bar{u}+6\bar{u}^2) \log \bar{u}
 \right. \nonumber\\
 &&\left. +u\bar{u}(2+13u\bar{u})\right\} \, ,\nonumber\\
 g_\pi(u)&=&1+g_2C^{\frac{1}{2}}_2(2u-1)+g_4C^{\frac{1}{2}}_4(2u-1)\, ,\nonumber\\
 B(u)&=&g_\pi(u)-\varphi_\pi(u)\, ,
\end{eqnarray}
where
\begin{eqnarray}
h_{00}&=&v_{00}=-\frac{\eta_4}{3} \, ,\nonumber\\
a_{10}&=&\frac{21}{8}\eta_4 \omega_4-\frac{9}{20}a_2 \, ,\nonumber\\
v_{10}&=&\frac{21}{8}\eta_4 \omega_4 \, ,\nonumber\\
h_{01}&=&\frac{7}{4}\eta_4\omega_4-\frac{3}{20}a_2 \, ,\nonumber\\
h_{10}&=&\frac{7}{2}\eta_4\omega_4+\frac{3}{20}a_2 \, ,\nonumber\\
g_2&=&1+\frac{18}{7}a_2+60\eta_3+\frac{20}{3}\eta_4 \, ,\nonumber\\
g_4&=&-\frac{9}{28}a_2-6\eta_3\omega_3 \, ,
\end{eqnarray}
$ C_2^{\frac{1}{2}}(\xi)$, $ C_4^{\frac{1}{2}}(\xi)$,
  $ C_2^{\frac{3}{2}}(\xi)$ and $ C_4^{\frac{3}{2}}(\xi)$ are Gegenbauer polynomials,
  $\eta_3=\frac{f_{3\pi}}{f_\pi}\frac{m_u+m_d}{m_\pi^2}$ and  $\rho^2={(m_u+m_d)^2\over m_\pi^2}$
 \cite{LCSR1,LCSR2,LCSR3,LCSR4,LCSR5,Belyaev94,PSLC1,PSLC2,PSLC3,PSLC4}.


\begin{thebibliography}{99}
\bibitem{Godfray} S. Godfray and J. Napolitano, Rev. Mod. Phys. {\bf 71 } (1999) 1411.

\bibitem{Review1} F. E. Close and N. A. Tornqvist, J. Phys.  {\bf G28} (2002) R249.

\bibitem{Review2} R. L. Jaffe, Phys. Rept. {\bf 409} (2005) 1.

\bibitem{Review3}  C. Amsler and N. A. Tornqvist, Phys. Rept. {\bf 389} (2004) 61.


\bibitem{4quark1} R. L. Jaffe, Phys. Rev. {\bf D15} (1977) 267, 281.

\bibitem{4quark2} R. L. Jaffe, Phys. Rev. {\bf D17} (1978) 1444.


\bibitem{KK1} J. D. Weinstein and N. Isgur, Phys. Rev. {\bf D41} (1990) 2236.

\bibitem{KK2} J. D. Weinstein and N. Isgur, Phys. Rev. {\bf D27} (1983) 588.


\bibitem{HDress1} N. A. Tornqvist, Z. Phys. {\bf C68} (1995) 647.

\bibitem{HDress2} M. Boglione and  M. R. Pennington, Phys. Rev. Lett. {\bf 79} (1997)
1998.

\bibitem{HDress3} M. Boglione and  M. R. Pennington, Phys.
Rev. {\bf D65} (2002) 114010.


\bibitem{Radiative1}  A. Bramon, G. Colangelo, P. J. Franzini and M. Greco, Phys.
Lett. {\bf B287} (1992) 263.

\bibitem{Radiative2} P. J. Franzini, W. Kim and J. Lee-Franzini, Phys. Lett. {\bf B287} (1992) 259.

\bibitem{Radiative3} G. Colangelo and P. J. Franzini, Phys. Lett. {\bf B289} (1992) 189.

\bibitem{Radiative4} N. N. Achasov, V. V. Gubin and E. P. Solodov, Phys. Rev. {\bf D55} (1997) 2672.

\bibitem{Radiative5} A. Bramon, A. Grau and G. Pancheri, Phys. Lett. {\bf B289} (1992) 97.

\bibitem{Radiative6} A. Bramon, R. Escribano, J. L. Lucio M, M. Napsuciale and G.
Pancheri, Eur. Phys. J. {\bf C26} (2002) 253.

\bibitem{Radiative7} E. Marco, S. Hirenzaki, E. Oset and H. Toki, Phys. Lett. {\bf B470} (1999) 20.


\bibitem{Wang23} Z. G. Wang and S. L. Wan, Phys. Rev. {\bf D73} (2006)
094020.

\bibitem{Wang24} Z. G. Wang, J. Phys. {\bf G34} (2007) 753.

\bibitem{Wang06} Z. G. Wang and S. L. Wan, Phys. Rev. {\bf D74} (2006) 014017.

\bibitem{Colangelo03} P. Colangelo and F. D. Fazio, Phys. Lett. {\bf B559} (2003)
49.

\bibitem{Wang04} Z. G. Wang, W. M. Yang and S. L. Wan,  Eur. Phys. J. {\bf C37}
(2004) 223.

\bibitem{Wang08} Z. G. Wang, Phys. Rev. {\bf D77} (2008) 054024.



\bibitem{eta-Yilmaz} A. Gokalp, Y. Sarac and O. Yilmaz, Mod. Phys. Lett. {\bf A19} (2004) 3011.


\bibitem{f0-sigma} A. Gokalp, Y. Sarac and O. Yilmaz, Phys. Lett. {\bf B609} (2005) 291.



\bibitem{LCSR1} I. I. Balitsky, V. M. Braun and A. V. Kolesnichenko, Nucl. Phys.
{\bf B312} (1989) 509.

\bibitem{LCSR2}  V. L. Chernyak and I. R. Zhitnitsky, Nucl.
Phys. {\bf B345} (1990) 137.

\bibitem{LCSR3} V. L. Chernyak and A. R. Zhitnitsky, Phys. Rept. {\bf 112} (1984)
173.

\bibitem{LCSR4}V. M. Braun and I. E. Filyanov, Z. Phys.  {\bf C44} (1989)
157.

\bibitem{LCSR5}V. M. Braun and I. E. Filyanov, Z. Phys. {\bf C48} (1990) 239.


\bibitem{SVZ791} M. A. Shifman, A. I. Vainshtein and V. I. Zakharov,
Nucl. Phys. {\bf B147} (1979) 385.

\bibitem{SVZ792} M. A. Shifman, A. I. Vainshtein and V. I. Zakharov,
Nucl. Phys. {\bf B147} (1979) 448.

\bibitem{Reinders85}  L. J. Reinders, H.
Rubinstein and S. Yazaki, Phys. Rept. {\bf 127} (1985) 1.



\bibitem{Belyaev94}
V. M. Belyaev, V. M. Braun, A. Khodjamirian and R. R\"uckl, Phys.
Rev. {\bf D51} (1995) 6177.

\bibitem{PSLC1} P. Ball, JHEP {\bf 9901} (1999) 010.

\bibitem{PSLC2} P. Ball and R. Zwicky, Phys. Lett. {\bf B633} (2006) 289.

 \bibitem{PSLC3} P. Ball and R. Zwicky, JHEP {\bf 0602} (2006) 034.

 \bibitem{PSLC4} P. Ball, V. M. Braun and  A. Lenz, JHEP {\bf 0605} (2006) 004.


\bibitem{decay-eta} L. Burakovsky and J. T. Goldman, Phys. Lett. {\bf B427} (1998) 361.

\bibitem{PDG} W. M. Yao  et al, J. Phys. {\bf G33} (2006) 1.


\bibitem{Schmedding1999-24} A. Schmedding and O. I. Yakovlev, Phys. Rev. {\bf D62} (2000) 116002.

\bibitem{Bakulev2001-24} A. P. Bakulev, S. V. Mikhailov and N. G. Stefanis, Phys. Lett. {\bf B508} (2001)
279.

\bibitem{Bakulev2002-24} A. P. Bakulev, S. V. Mikhailov and N. G. Stefanis, Phys. Rev. {\bf D67} (2003) 074012.

\bibitem{Bakulev2003-24} A. P. Bakulev, S. V. Mikhailov  and N. G. Stefanis, Phys. Lett. {\bf B578} (2004) 91.

\bibitem{Bakulev2006-24} A. P. Bakulev and A. V. Pimikov, Acta. Phys. Polon. {\bf B37} (2006) 3627.

\bibitem{Bakulev2006-24-2} A. P. Bakulev, hep-ph/0611139.


\bibitem{4Coupling} T. V. Brito, F. S. Navarra, M. Nielsen and M. E. Bracco,  Phys. Lett. {\bf B608} (2005) 69.



\bibitem{Achasov89} N. N. Achasov and V. N. Ivanchenko, Nucl. Phys. {\bf
B315} (1989) 465.

\bibitem{Achasov97}  N. N. Achasov and V. V. Gubin, Phys. Rev. {\bf D56} (1997)
4084.

\bibitem{SND1}  M. N. Achasov  et al, Phys. Lett. {\bf B485} (2000) 349.


\bibitem{SND2} M. N. Achasov et al, Phys. Lett. {\bf B479} (2000) 53.


\bibitem{KLOE1}  A. Aloisio et al, Phys. Lett. {\bf B536} (2002)
209.

\bibitem{KLOE2} A. Aloisio et al,  Phys. Lett. {\bf B537} (2002)
21.

\bibitem{KLOE07} F. Ambrosino et al, arXiv:0707.4609.

\end{thebibliography}
\end{document}